\documentclass[11pt]{article}
\usepackage{jheppub}
\usepackage{amsmath,amssymb,amsfonts,amsthm}
\usepackage{bm}
\usepackage{graphicx}
\usepackage[dvipsnames]{xcolor}
\usepackage{hyperref}

\makeatletter
\def\@fpheader{\relax}
\makeatother

\renewcommand{\footnoterule}{\vfill\kern -3pt \hrule width 0.4\columnwidth \kern 2.6pt} 

\DeclareMathOperator{\Tr}{Tr}
\DeclareMathOperator{\Kh}{Kh}
\DeclareMathOperator{\Vol}{Vol}

\newcommand\vev[1]{\langle #1 \rangle}
\newcommand\be{\begin{equation}}
\newcommand\ee{\end{equation}}
\newcommand\bea{\begin{eqnarray}}
\newcommand\eea{\end{eqnarray}}

\newcommand\eref[1]{(\ref{#1})}
\newcommand\bc{\begin{center}}
\newcommand\ec{\end{center}}

\numberwithin{equation}{section} 
\allowdisplaybreaks

\title{
(K)not machine learning
}

\author[a]{Jessica Craven}
\author[b]{\!, Mark Hughes}
\author[a]{\!, Vishnu Jejjala}
\author[c]{\!, Arjun Kar}

\affiliation[\,a]{Mandelstam Institute for Theoretical Physics, School of Physics, NITheCS, and CoE-MaSS,\\ University of the Witwatersrand, 1 Jan Smuts Avenue, Johannesburg, WITS 2050, South Africa}
\affiliation[\,b]{Department of Mathematics, Brigham Young University,\\ 275 TMCB, Provo, UT 84602, USA}
\affiliation[\,c]{Department of Physics and Astronomy, University of British Columbia,\\ 6224 Agricultural Road, Vancouver, BC V6T 1Z1, Canada}

\emailAdd{jessica.craven1@students.wits.ac.za}
\emailAdd{hughes@mathematics.byu.edu}
\emailAdd{vishnu@neo.phys.wits.ac.za}
\emailAdd{arjunkar@phas.ubc.ca}

\abstract{
We review recent efforts to machine learn relations between knot invariants.
Because these knot invariants have meaning in physics, we explore aspects of Chern--Simons theory and higher dimensional gauge theories.
The goal of this work is to translate numerical experiments with Big Data to new analytic results.
}

\begin{document}

\maketitle
\parskip=10pt

\section{Introduction}
In numerous settings, machine learning successfully identifies associations between features in large data sets.
Neural networks, which are often used to model these relationships, are universal approximators~\cite{Cybenko1989,Hornik1991}.
They are also black boxes.
Even when the predictions are highly accurate, we typically do not know how the machine learns.
In certain examples in algebraic geometry, the way the training time scales with the size of the input is sufficient to surmise that the machine makes its prediction in a different manner than how a human would perform the calculation~\cite{Bull:2018uow,Bull:2019cij}.
This suggests that there are better ways to calculate~\cite{Brodie:2019dfx,Brodie:2019pnz}.
As theoretical physicists and mathematicians, we aim to obtain simple analytic expressions that are descriptive of a system of interest and to refine our methods of computation.
We seek to promote machine learning to a discovery tool.

One advantage of the large data sets in mathematics and formal theory is that they are clean.
There is no measurement error or noise infecting the data.
Looking at such exact data from a phenomenological angle has historically yielded new insights.
Mirror symmetry in Calabi--Yau manifolds was an experimental observation about string compactification and conformal field theories~\cite{Dixon:1987bg,Lerche:1989uy,Candelas:1989hd,Greene:1990ud} before it became a mathematical fact~\cite{givental1996,Lian:1997fkq,Lian:1999rp}.
There are half a billion four dimensional reflexive polytopes with which to explore this relationship further~\cite{Kreuzer:2000xy}.
Similarly large data sets exist in knot theory.
As with Calabi--Yau manifolds, the knot theory data sets can be approached on purely mathematical grounds or in terms of theoretical physics.

A knot is a circle $S^1$ embedded in a three manifold $\mathcal{M}$, which for simplicity, we may take to be $S^3$.
Because the representation of a knot is not unique, we must use topological data to distinguish one knot from another.
These topological invariants are tabulated in several databases~\cite{KnotAtlas,knotinfo} or can be calculated from code~\cite{KnotTheory,KnotJob}.
They can also be constructed in quantum field theory.
The existing knot invariants are not independent.
Some relations between them are known, while others can be experimentally found and mathematically proved, sometimes using artificial intelligence~\cite{davies2021advancing,davies2021signature}.

In this article, which summarizes the results of~\cite{Jejjala:2019kio,Craven:2020bdz,Craven:2021ckk}, we explore new relations between knot invariants.
In Section~\ref{sec:dp}, we introduce the relevant knot invariants.
In Section~\ref{sec:ml}, we apply machine learning to the data.
In Section~\ref{sec:p}, we outline targets for future work in this area.

\section{Dramatis person\ae}\label{sec:dp}
The \textit{Jones polynomial} is a Laurent polynomial computed using the Kauffman bracket~\cite{jones85}.
While the mathematical definition is intrinsically two dimensional, Witten established that this polynomial knot invariant has a physical meaning as the unknot normalized vacuum expectation value of a Wilson loop operator in three dimensional $SU(2)$ Chern--Simons theory~\cite{Witten:1988hf}:
\bea
J_n(K;q) &=& \frac{\int_\mathcal{U} [DA]\ U_n(K) e^{i S_\text{CS}}}{\int_\mathcal{U} [DA]\ U_n(\bigcirc) e^{i S_\text{CS}}} = \frac{\vev{U_n(K)}}{\vev{U_n(\bigcirc)}} ~, \label{eq:jones} \\\label{eq:cs}
S_\text{CS} &=& \frac{k}{4\pi} \int_\mathcal{M} \Tr\, \big(A\wedge dA + \frac23 A\wedge A\wedge A \big) ~. 
\eea
The path integral is taken over $SU(2)$ connections modulo gauge transformations.
To ensure that the action is gauge invariant, the coupling $k$, or Chern--Simons level, is integer quantized.
The Wilson loops are
\be
U_R(\gamma) = \Tr_R\, \mathcal{P} \exp \big( i\oint_\gamma A \big) \label{eq:wilson}
\ee
and denote the trace in the representation $R$ of the path ordered exponential of the holonomy of the gauge connection along some curve $\gamma\subset \mathcal{M}$, which in~\eref{eq:jones} we take to be the knot $K$.
The ordinary Jones polynomial $J_2(K;q)$ corresponds to an evaluation of~\eref{eq:jones} at $q=e^\frac{2\pi i}{k+2}$, with $n=2$ denoting the fundamental representation of $SU(2)$.
Because it is a topological invariant, the expression is independent of the metric on $\mathcal{M}$.
The Jones polynomial does not uniquely identify a knot.
In fact, it is not known whether the Jones polynomial even uniquely identifies the unknot.

The coefficients in the Jones polynomial are integer valued.
This is clear from the mathematical construction but surprising from the definition using Chern--Simons theory.
This is also evident through the work of Khovanov, who defined bigraded groups $\mathcal{H}(K)$ for any knot $K$~\cite{khovanov2000,Bar_Natan_2002}.
The information in Khovanov's homology theory can be captured by the \textit{Khovanov polynomial}:
\be
\Kh(K; q, t) = \sum_{m,n} \dim \mathcal{H}^{m,n}(K) t^m q^n ~, \label{eq:kh}
\ee
where $m$ and $n$ are, respectively, the homological grading and the quantum grading.
Because the coefficients are the dimensions of certain homology groups, they are integer valued.
A specialization of the Khovanov polynomial recovers the Jones polynomial:
\be
J_2(K;q^2) = \frac{\Kh(K; q, -1)}{q+q^{-1}} ~.
\ee
As the denominator is the Khovanov polynomial of the unknot, this expression for the Jones polynomial is appropriately normalized.
While Khovanov's definition of $\mathcal{H}^{m,n}(K)$ in~\eref{eq:kh} is also two dimensional, it is conjectured to be related to quantum field theories in higher dimensions.
From a four dimensional perspective, there is a super-Yang--Mills path integral that counts classical supersymmetric solutions weighted by $(-1)^F q^I$, where the homological grading is identified with the fermion number and the quantum grading with the instanton number~\cite{Gaiotto:2011nm}.
Going up a dimension allows for a Hilbert space interpretation of $\mathcal{H}(K)$.
In particular, the space $\mathcal{H}(K)$ is given by the $Q$-cohomology in the Hilbert space of a five dimensional super-Yang--Mills theory, with a bigrading given by a $U(1)\times U(1)$ symmetry that is generated by fermion and instanton number operators~\cite{Witten:2011zz}.
This five dimensional theory is not ultraviolet complete, and there is as well a six dimensional description with $\mathcal{H}(K)$ the cohomology of a supercharge $Q$ in the $(0,2)$ M$5$-brane theory on a Cauchy slice with surface operator $\Sigma_K$ whose topology is set by the knot~\cite{Witten:2011zz}.

Many knots --- and nearly every knot at small crossing number --- are hyperbolic.
For such knots, when we excise an $\epsilon$-size tubular neighborhood around the knot, the knot complement $S^3 \setminus K$ admits a unique constant negative curvature metric~\cite{mostow1968quasi}.
Thurston showed that the \textit{hyperbolic volume} computed using this metric is a knot invariant that can be conveniently calculated from a tetrahedral decomposition~\cite{thurston}.
Complexifying the gauge group of the Chern--Simons theory from $SU(2)$ to $SL(2,\mathbb{C})$, the volume appears as the saddle point in the partition function~\cite{Witten:2010cx}
\be
\mathcal{Z}(\mathcal{M}) = \int_{\mathcal{U}_\mathbb{C}} [D\mathcal{A}][D\bar{\mathcal{A}}] \exp \left( \frac{it}{2} \mathcal W(\mathcal{A}) + \frac{i\tilde{t}}{2} W(\bar{\mathcal{A}}) \right) ~.
\ee
The critical point corresponds to a flat $SL(2,\mathbb{C})$ connection $\mathcal{A}_+$ known as the geometric conjugate connection:
\be
W(\mathcal{A}_+) = -\frac{i}{2\pi}\Vol(S^3\setminus K) + \pi \text{CS}(S^3\setminus K) ~. \label{eq:gcc}
\ee
(The second term in~\eref{eq:gcc} is proportional to the Chern--Simons invariant of the knot.)
The complexification serves as a physical motivation for the volume conjecture~\cite{Kashaev1997,Murakami2001,murakami2002kashaev,Gukov:2003na}, which states
\be
\Vol(S^3\setminus K) = \lim_{n\to\infty} \frac{2\pi \log |J_n(K; \omega_n)|}{n} ~, \qquad \omega_n = e^\frac{2\pi i}{n} ~.
\ee
This is implicitly at large-$k$ as well.

Two other mathematical invariants are the \textit{Rasmussen $s$-invariant} and the \textit{smooth slice genus $g$}.
The $s$-invariant is defined using the Lee spectral sequence~\cite{lee2008khovanov} which collapses $\mathcal{H}(K)$ to $\mathbb{Q}\oplus \mathbb{Q}$ with instanton grading $s\pm 1$ and $s\in 2\mathbb{Z}$~\cite{rasmussen2010khovanov}.
The slice genus of a knot $K$ is the least integer $g$ for which there is a smooth orientable surface $\Sigma\subset B^4$ with genus $g$ such that $K = \partial \Sigma \subset S^3$.
The slice genus constrains the $s$-invariant~\cite{rasmussen2010khovanov}:
\be
|s| \le 2g ~. \label{eq:bound}
\ee
These are four dimensional invariants.
To date, despite notable efforts~\cite{kronheimer2013,Gukov:2015gmm}, there is no simple gauge theoretic interpretation of these quantities.

\section{Learning from polynomial invariants}\label{sec:ml}
Fully connected neural networks will be our principal machine learning tool.
For our purposes, two hidden layer networks with $\mathcal{O}(100)$ neurons per layer are sufficient to extract relationships between knot invariants.
As the results are not especially sensitive to the architecture of the network or the precise non-linearities introduced, the results could be interpretable.
Indeed, this is our objective.

\begin{figure}[h]
    \centering
    \includegraphics[scale=.1]{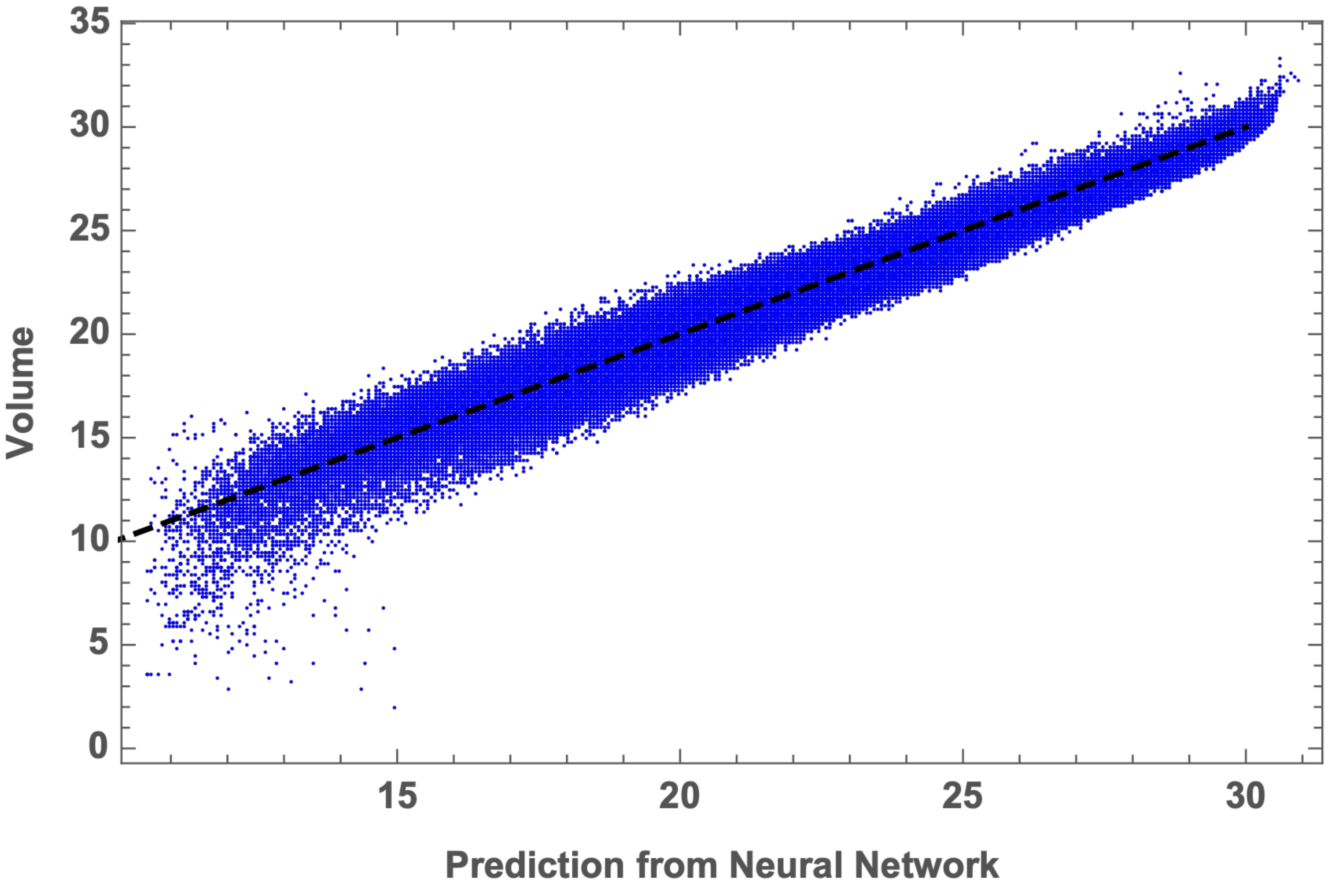}
    \caption{\small\textsf{Prediction of hyperbolic volume from Jones polynomial using knots up to $15$ crossings with training fraction $0.1$.
    This figure appears in~\cite{Jejjala:2019kio}.}}
    \label{fig:volume}
\end{figure}

There are $313,209$ hyperbolic knots up to fifteen crossings.
The Jones polynomials have up to sixteen coefficients.
Including the minimum and maximum degree, there are eighteen input features.
Using only $10\%$ of the data for training, a neural network accomplishes $2.45\pm 0.10\%$ mean relative error in its predictions of the volume.
Figure~\ref{fig:volume} plots the true volume vs.\ the neural network prediction.
When knots with the same Jones polynomial have knot complements with different volumes, the volumes differ by about $2.83\%$.
Thus, the neural network predicts the volume near the theoretical upper limit of performance.

The Jones polynomial is defined through Chern--Simons theory.
It is a polynomial invariant that is intrinsically quantum in character.
In contrast, the volume of the knot complement is a classical knot invariant.
Its connection to the quantum field theory arises through the volume conjecture only in the large-$n$ limit.
Nevertheless, one predicts the other.
Why?

\begin{figure}[h]
    \centering
    \includegraphics[scale=.33]{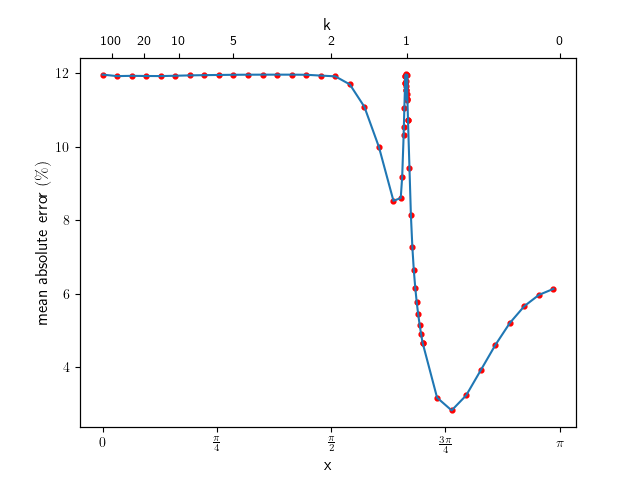}
    \caption{\small\textsf{Regression to predict volume based on evaluation of the Jones polynomial at different phases.
    This figure appears in~\cite{Craven:2020bdz}.}}
    \label{fig:phase}
\end{figure}

It turns out that the degrees of the polynomials are superfluous data for an accurate prediction.
This suggests that the evaluation of the Jones polynomial at particular phases can also predict the volume.
By feeding a neural network a set of these evaluations, we use layer-wise relevance propagation~\cite{montavon2019layer} to assign a relevance score to each neuron and to each input feature.
The evaluation of the Jones polynomial at the phase $e^\frac{3\pi i}{4}$ has the largest relevance score.
From regression, we find that the formula
\be
\Vol(S^3\setminus K) = 6.20\log(|J_2(K;e^\frac{3\pi i}{4})| + 6.77) - 0.94  \label{eq:bestfit}
\ee
gives an error of only $2.86\%$ on $1,701,903$ hyperbolic knots up to sixteen crossings.
Equating the phase to $q=e^\frac{2\pi i}{k+2}$, this corresponds to a fractional Chern--Simons level $k=\frac23$.
In Figure~\ref{fig:phase}, we plot the best fit of the form in~\eref{eq:bestfit} against the argument of the phase.

This is interpretable in terms of the analytically continued Chern--Simons theory~\cite{Witten:2010cx}.
The approximation formula works well for the levels $k$ for which the geometric conjugate connection $\mathcal{A}_+$ makes a contribution to the Chern--Simons path integral, and the accuracy of the prediction increases with the fraction of the knots in the data set that receive such a contribution.
The plateau at $k>2$ corresponds to latent correlations in the data set, \textit{viz.}, equating the volume of every knot to the mean volume of knots in the data set gives a $12\%$ error.
The minimum occurs near $k=\frac23$.
At integer values of $k$, the path integral receives contributions only from $SU(2)$ valued critical points, meaning that we lose information from the geometric conjugate connection.
This explains the spike at $k=1$.

Working with a data set comprised of $535,239$ knots for which we know the Khovanov and Jones polynomials, the $s$-invariant, and for which we know the slice genus in $438,295$ cases, we find that the Khovanov polynomial correctly predicts both the Rasmussen $s$-invariant and the slice genus more than $98\%$ of the time.
The Jones polynomial, respectively, predicts the Rasmussen $s$-invariant and the slice genus correctly $95\%$ and $97\%$ of the time.
There is no known way of calculating the slice genus from the Khovanov polynomial.
The success of the Khovanov polynomial in predicting the $s$-invariant is somewhat less surprising.
The knight move conjecture~\cite{Bar_Natan_2002}, which is false~\cite{manolescu2018knight}, is satisfied by all the knots in our data set.
It states
\be
\Kh(K;q,-q^{-4}) = q^s(q+q^{-1}) ~.
\ee
An evaluation of the Khovanov polynomial at $t=-q^{-4}$ is successful at predicting the $s$-invariant $99.77\%$ of the time.
Surprisingly, an evaluation at $t=-q^{-2}$ is successful $99.88$\% of the time.
In fact, the polynomials often have particular forms:
\bea
\Kh(K;q,-q^{-2}) &=& \big( \frac{s}{2}+1 \big) q^{s-1} - \big( \frac{s}{2}-1 \big) q^{s+1} ~, \\
\Kh(K;q,-q^{-2}) &=& -q^{s-1} + \big( \frac{s}{2}+5 \big) q^{s+1} - \big( \frac{s}{2}+2 \big) q^{s+3} ~. 
\eea
The exponents being dependent on $s$ is a mathematical renormalization of the Khovanov polynomial.
Why the $s$-invariant appears in the coefficients is theoretically unmotivated.
However, the experiments suggest that the $s$-invariant is encoded in the Khovanov polynomial in more than one way.
The ability of the Jones polynomial to predict the Rasmussen $s$-invariant and the slice genus is a mystery.
The lower bound in the inequality~\eref{eq:bound} is saturated by $414,615$ knots, so it is also not clear from numerical experiments whether the network is learning $s$ or $g$.

\section{Prospectus}\label{sec:p}
Ever since~\cite{hughes2016neural}, knot invariants have been a subject of machine learning.
Besides the papers we have reviewed here, other machine learning investigations on related data sets appear in~\cite{levitt2019big,Gukov:2020qaj,pawel2021knot,davies2021advancing}.
Our efforts point to deep relations in mathematics and physics.
Perhaps there are algebraic formulas for the Rasmussen $s$-invariant and the slice genus.
The knot invariants we examine are realized in quantum field theories in different dimensions.
Understanding how these quantum field theories are related is work in progress.
As well,~\cite{freedman2010man} proposed a strategy for finding counterexamples to the smooth four dimensional Poincar\'e conjecture by looking for slice genus zero knots with certain properties.
A neural network that accurately discriminates $g=0$ knots may prove useful in this endeavor.
Man and machine continue thinking about these problems.

\section*{Acknowledgments}
We thank Onkar Parrikar for collaboration on~\cite{Jejjala:2019kio}.
The work of JC and VJ is based on research supported in part by the South African Research Chairs Initiative of the National Research Foundation, grant number 78554.
VJ is additionally supported by a Simons Foundation Mathematics and Physical Sciences Targeted Grant, 509116.
AK is supported by the Simons Foundation through the It from Qubit Collaboration.
These proceedings are based on a talk by VJ at the Nankai Symposium on Mathematical Dialogues held at the Chern Institute of Mathematics in August 2021.
We are grateful to the organizers of this meeting.

{\small
\bibliographystyle{JHEP}
\bibliography{knots}
}

\end{document}